# English dictionaries, gold and silver standard corpora for biomedical natural language processing related to SARS-CoV-2 and COVID-19

***Salma Kazemi Rashed[1], Rafsan Ahmed[1], Johan Frid[2], Sonja Aits[1] ****

[1] Cell Death, Lysosomes and Artificial Intelligence Group, Department of Experimental Medical Science, Faculty of Medicine, Lund University, Lund, Sweden
[2] Humanities Laboratory, Lund University, Lund, Sweden

[*] For correspondence contact: sonja.aits@med.lu.se

**Abstract**

***Background***

Automated information extraction with natural language processing (NLP) tools is required to gain systematic insights from the large number of COVID-19 publications, reports and social media posts, which far exceed human processing capabilities.

***Results***

Here we present an NLP toolbox comprising COVID-19-related dictionaries and annotated corpora in English as well as useful code and workflows for their update and use. The dictionaries contain terms referring to the COVID-19 disease, the SARS-CoV-2 virus, its variants and common mutations, respectively. They were used together with the EasyNER NLP tool to extract and annotate all 764 398 abstracts in the CORD-19 dataset, creating a very large silver standard corpus (named Lund-Annotated-CORD-19 corpus). This was complemented with a small gold standard corpus consisting of PubMed abstracts manually annotated for key entity classes such as disease, virus, symptom, protein/gene, cell type, chemical and species terms.
The toolbox can support various text analysis tasks related to COVID-19 such as named entity recognition and co-mention analysis. A preliminary version of the toolbox, which was released early in the pandemic, was for example already used to create a COVID-19 knowledge graph and study the evolution and variation of COVID-19-related terminology. In addition, the toolbox can be applied in the development of other NLP tools, for example to train and evaluate large language models.
Analysis of the Lund-Annotated-CORD-19 corpus, which represents a large section of the coronavirus-related literature published until 2022, can provide both linguistic and medical insights. We observed matches for hundreds of SARS-CoV-2 and COVID-19 synonyms, indicating a high degree of term variability, which has also been reported for other datasets. Terms referring to the disease were the most frequent by far, followed by terms referring to the virus. We also found thousands of mentions of variants and mutations. However, most of these referred to a small group of highly studied variants and mutations, reflecting research biases and revealing understudied aspects of the virus.

***Conclusions***

The presented toolbox has a broad variety of NLP applications related to COVID-19. It is freely available on GitHub (on https://github.com/Aitslab/Covid19) and Zenodo (https://doi.org/10.5281/zenodo.15395348).

## Introduction

The analysis of various types of text related SARS-CoV-2/COVID-19 has many applications. For example, researchers may use it to guide their work and public health authorities may rely on it to monitor outbreaks, guide policies or track misinformation. It can also be used for many other purposes outside of the medical domain, for example by scientists studying the social impact of the pandemic or the linguistic development that accompanied it, by libraries/archives or by software developers working on directed search tools. In many contexts, this type of analysis is the most meaningful when conducted with large text datasets which requires computational rather than manual processing, which is referred to as natural language processing (NLP). In many types of text analysis, named entity recognition (NER), which refers to the identification of relevant keywords and phrases is often a key step. This involves the detection of many synonymous terms which describe the same entity.

NER can be especially challenging with newly emerging diseases such as COVID-19 where a large variation in terminology can occur. This is because no official name had been defined during the early phases of the outbreak [1] and because such a large and diverse global community was involved in the writing of COVID-19-related texts. Many authors thus used a wide range of descriptive terms such as "Wuhan seafood market pneumonia" [2]. Any analysis using only the official name as search term would thus be extremely flawed. Instead, NER can be performed using dictionaries, essentially long lists of keywords and their various synonyms, which can be created by manual curation and/or computational approaches. Alternatively, machine learning models can be used, which are normally trained on annotated corpora in which keywords have been labelled manually by experts (gold standard) or automatically (silver standard) [3]. Dictionary and model-based NER methods can also be combined as the approaches are complimentary.

Here we present a toolbox for NLP related to SARS-CoV-2 and COVID-19. It includes English dictionaries of synonymous terms for four key entity types as well as an annotated gold and silver standard corpora and associated code, workflows and usage instructions. A first version was released at an early stage of the pandemic with the preprint of this article. It has already been used for various applications, for example to generate COVID-19 knowledge graphs [4], cluster scientific articles [5] and analyse the variation and evolution of COVID-19 terminology [6]. We have now substantially expanded the toolbox. It is freely available on GitHub (on https://github.com/Aitslab/Covid19) and Zenodo (https://doi.org/10.5281/10.5281/zenodo.15395348) [7] so it can be used by medical professionals, data scientists and others interested in analysing COVID-19-related texts.

## Methods

### *Generation of SARS-CoV-2-related dictionaries*

To generate the dictionaries with terms referring to SARS-CoV-2 (virus dictionary, Supplemental file 1) or COVID-19 (disease dictionary, Supplemental file 2) synonyms and biomedical identifiers were collected by reviewing a variety of databases and text sources including: NCBI Taxonomy database [8], Wikidata, the International Classification of Diseases - v10 and v11[9, 10], Disease Ontology [11], Medical Subject Headings (MeSH) [12], medical research literature, twitter feeds, and newspaper websites. Terms lists created by the Leaman and Lu group were also added [6, 13], but excluding terms containing commas, full-stops or colons (e.g. "COVID-19, infection"), spelling error variants (e.g. "CVOID-19"), highly ambiguous (e.g. "CI), non-sensical terms (e.g. "novel coronavirus Coronavirus Disease") and terms referring to a patient characteristic (e.g. COVID-19 infected), rather than the disease itself. Ambiguous abbreviations (e.g. NCP - novel coronavirus pneumonia) were removed since they can be detected through abbreviation resolution. Variants for virus names and disease names were generated as follows:
Virus names:

- Adding '2019', '2019novel', '2019new', '2019 novel', '2019 new' as prefixes or 2019 as suffix (only if '19' was not in the virus name)
- Interchanging corona virus and coronavirus
- Removing virus names containing the same word twice or containing both 'new' and 'novel'
- Interchanging Wuhan and Hubei

Disease names:

- interchanging corona virus and coronavirus
- pairing the terms of the virus list with generic terms indicating the disease, e.g. illness, disorder, disease, pneumonia
- interchanging the terms disease, disorder, syndrome, pneumonia, infection
- adding the adjectives acute, severe, and respiratory alone or in combination
- removing disease names containing the same word twice or containing both 'new' and 'novel'

Duplicate dictionary entries as well as automatically generated entries with duplicate or semantically similar words (i.e. containing both “novel” and “new”, both “Hubei” and “Wuhan” or both “corona virus” and “coronavirus”) were removed.
To generate a dictionary of SARS-CoV-2 variant names (variant dictionary, Supplemental file 3), variant (lineage) and clade names were collected on 2022-01-13 from the WHO [14], GSAID [15], Nextstrain [16, 17] and Pango [18]. Updated taxon, lineage and clade names were collected from the Nextstrain (same link as before) and Pango [19, 20] on 2025-03-19 and added to the variant dictionary. This also included names referring to specific samples (e.g. “USA/NY-Wadsworth-21016210-01/2021”). For clade names, we produced additional terms with the suffix or prefix “clade” and for the lineage terms, we produced additional terms with the suffixes and prefixes “lineage” and “variant”.
A dictionary of common protein mutations in SARS-CoV-2 (mutation dictionary, Supplemental file 4) was generated from mutations listed in two recent reviews [21, 22]. Simple lexical variations (e.g. removal of blank, replacement of deletion L245 with ΔL245) were also included.
To minimize the size of all dictionaries, hyphens were replaced with blanks (“ “), letters changed to lower case, and plural forms were resolved to singular (e.g. “infections” replaced with “infection”) where possible.
Scripts used to create the dictionaries can be found in Supplemental file 5 and further usage instructions are provided in the GitHub repository accompanying this article (https://github.com/Aitslab/Covid19).

### *Production and analysis of a CORD-19-based silver standard corpus*

To produce the Lund-Annotated-CORD-19 corpus (Supplemental file 6), titles and abstracts from the CORD-19 database [23], a large collection of research articles related to COVID-19, were annotated using all four dictionaries and EasyNER, version 2.0.0 [24, 25]. The final release of the CORD-19 database [23, 26], from June 2, 2022, was obtained and title and abstracts extracted from the *metadata.csv* file using the EasyNER *cord-loader* module, producing a text collection in JSON format. Abstract texts (but not titles) were then split into sentences using the EasyNER *splitter* module with the *en_core_web_sm-3.8.0* spaCy model. The EasyNER *NER* script “ner_spacy.py” was replaced with a version which implemented removal of hyphens from the text to match the dictionaries (Supplemental file 5). Sentences were then annotated by using the four dictionaries (Supplemental files 1-4) in the modified EasyNER *NER* module. For tokenization in this module, we used the same spaCy model as for sentence splitting. The frequencies of the identified terms were analysed using the EasyNER *analysis* module and a free-standing script which aggregated counts of terms that only differed in casing (e.g. COVID19 and Covid19). Randomly selected sentences containing the most common terms were inspected manually to identify common errors. After this, variant annotations were replaced with annotations produced with a modified variant dictionary from which terms containing less than three characters had been removed (Supplemental file 7), creating the final version of the Lund-Annotated-CORD19 corpus. However, due to CORD-19 license limitations, only part of the Lund-Annotated-CORD19 corpus (197 905 abstracts) can be released together with instructions for reproducing the full corpus (Supplemental file 6).

***Production and analysis of a COVID-19-related gold standard corpus***

Using BioQRator {Kwon, 2013 #1543, English abstracts related to COVID-19/SARS-CoV-2 and published between Dec 2019 and March 6 2020 were identified on PubMed with the following search query: (((COVID-19 OR SARS-CoV-2 OR (Wuhan AND virus) OR 2019-nCoV OR (Wuhan AND pneumonia)) AND English[lang])) AND ("2019/12"[Date - Create] : "2020"[Date - Create]). From the abstracts loaded into BioQRator without error, 10 were randomly selected and subsequently annotated for the following concepts to generate the "Lund-COVID-19" gold standard corpus:

- Virus_SARS-CoV-2: for terms representing SARS-CoV-2, including generic terms when they referred to this specific virus-based on the context (e.g. "the virus")
- Virus_other: for terms representing a specific virus other than SARS-CoV-2 (e.g. MERS, SARS-CoV)
- Virus_family: for terms representing more than one virus (e.g. coronaviruses) or all viruses; for this concept, unique identifiers from the NCBI taxonomy database were also indicated
- Cell: for terms describing specific cell types, e.g. "mast cells"
- Protein: for terms representing specific proteins or genes, but not a protein family (e.g.IL-2 is annotated under this concept, but not protease, IL-1 family members orinterleukins); for this concept, unique identifiers from the UniProt database were also indicated
- Disease_COVID-19: for terms representing COVID-19
- Disease_other: for terms representing a disease other than COVID-19 (e.g. Zika virus infection) or general terms for disease (e.g. infection)
- Symptom: for terms representing disease symptoms (e.g. fever, cough)."Pneumonia" was annotated as Disease_other or Symptom depending on the context.
- Species_human: for terms representing humans (e.g. "man", "patient")
- Species_other: for terms representing other species, including terms referring to groups of species (e.g. "mammals")

Abbreviations were annotated independently unless they were nested inside an expression, e.g. in the expression novel coronavirus (2019-nCoV) both "novel coronavirus" and "2019-nCoV" were annotated as Virus_SARS-CoV-2. Annotations were counted manually to obtain statistics.

The corpus was exported from BioQRator as a csv and BioC xml file and converted to BioC json format (Supplemental files 8-10) using the script ofthe BioC-JSON tool from the NLM/NCBI BioNLP Research Group {Comeau, 2013 #1321}[27] with minor changes (Supplemental file 5).

## Results

### *Production of large SARS-CoV-2/COVID-19-related dictionaries by manual term collection and computational augmentation*

Dictionaries can be used for NER with exact or fuzzy matching and can also be combined with model-based NER. For our toolbox, we generated four different dictionaries: 1. a dictionary of terms referring to SARS-CoV-2 (virus dictionary), 2. a dictionary of terms referring to COVID-19 (disease dictionary), 3. a dictionary of terms referring to SARS-CoV-2 variants (variant dictionary) and 4. a dictionary of terms referring to common SARS-CoV-2 mutations (mutation dictionary).

For the virus and disease dictionaries, terms were first collected by browsing through a variety of biomedical texts and other public text sources such as news reports and social media posts. For completeness, even biomedical identifiers, colloquial and derogatory terms (e.g. "China virus" and "kung flu") as well as terms which can match other biomedical entities (e.g. "coronavirus") were included. In contrast, highly ambiguous abbreviations (e.g. "CI", "CV") were excluded as these would result in many false-positive matches. The manually collected terms were expanded by computationally generating variations that account for common differences in spelling (e.g. "coronavirus" and "corona virus"), common pre-fixes (e.g. "2019") or interchangeably used terms ("infection" and "disease"). Duplicates and non-sensical expressions, which contained the same or similar words twice, were excluded. To minimize dictionary size, we removed hyphens, capitalization and plural forms. The first version of the dictionaries, published with the first pre-print version of this article, contained 215 SARS-CoV-2 virus synonyms and 12915 COVID-19 disease synonyms. The current versions are much larger, with 867 virus terms (Supplemental file 1) and 89 938 disease terms (Supplemental file 2).

We also created dictionaries for variant and lineage terms, including those that indicate specific samples, by extracting terms from relevant databases. In addition, we produced a dictionary with the most common mutations, which were extracted from relevant review articles. After similar computational postprocessing as for the disease and virus dictionaries, we initially obtained 2 633 758 variant terms (Supplemental file 3). A modified version containing only variant terms with more than 2 characters retained 1 133 377 terms (Supplemental file 7). The mutation dictionary consisted of 113 terms (Supplemental file 4). As additional terms will emerge over time, especially for the variants and mutations, our toolbox includes the scripts used to produce the dictionaries, as well as usage instructions (Supplemental file 5), so that end users can update the dictionaries when needed.

### *Production and analysis of a large, annotated silver standard corpus from the CORD-19 dataset*

Computationally annotated silver standard corpora can be used to train and benchmark NER tools, and it is possible to produce them at a much larger scale than corpora with manual annotations. To generate a large COVID-19-related silver standard corpus, we annotated all abstracts from the CORD-19 corpus [23], a large collection of coronavirus-related articles published until June 2022, with our dictionaries. For this, we used EasyNER, a customizable text mining tool developed in our group [24], which can extract and pre-process texts, perform NER with user-defined dictionaries/models, merge and analyse the results.

From processing the CORD-19 metadata file with EasyNER we obtained 764 398 abstracts with at least one sentence, which were then split into sentences, tokenized and subjected to NER with the four dictionaries (virus, disease, variants and mutations). Hyphen removal was added to the standard text pre-processing to ensure matching with our dictionaries and all individual NER files were merged to create the final Lund-Annotated-CORD-19 corpus in JSON format (Supplemental file 6).

A Top 50 SARS-CoV-2 terms

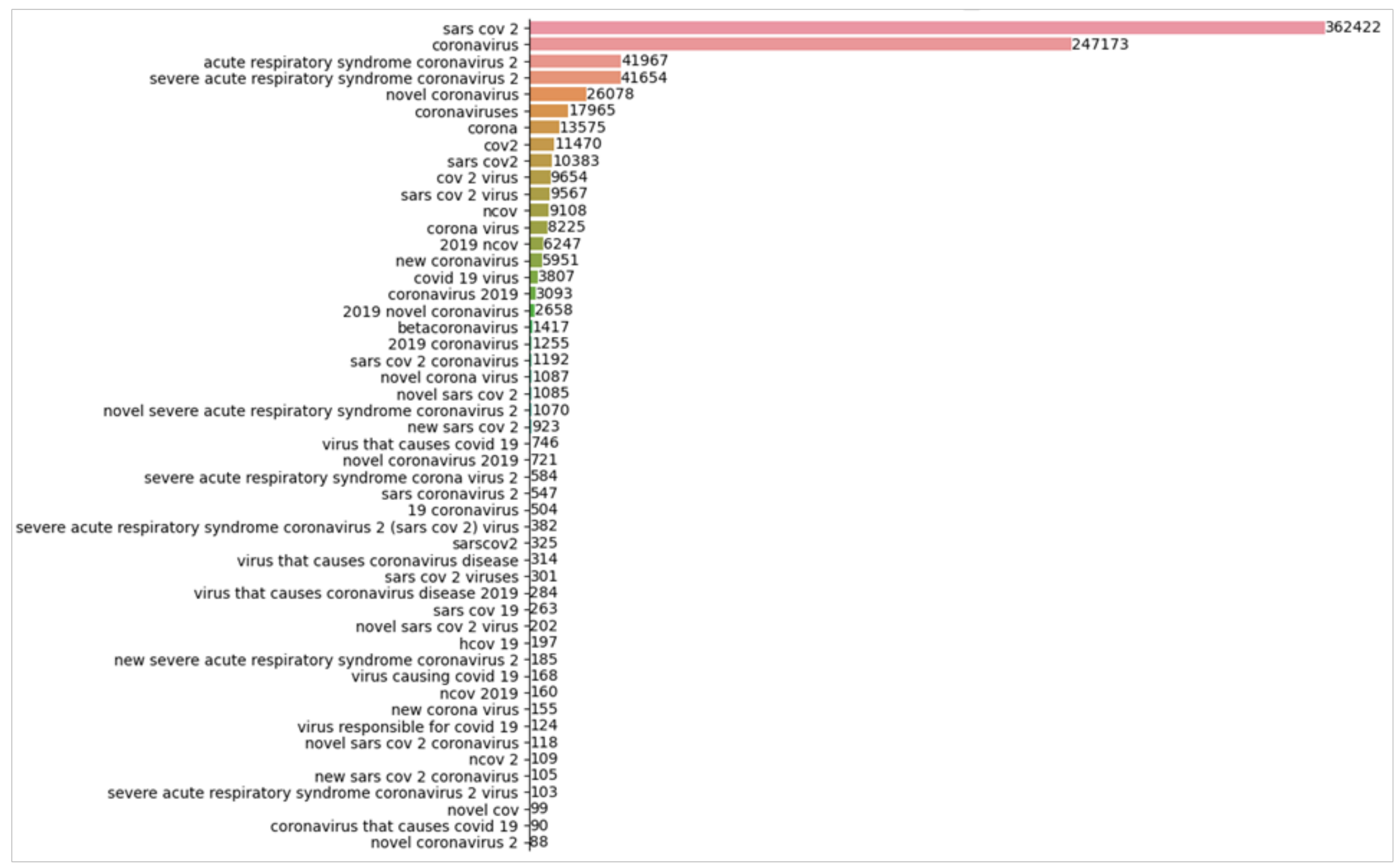

B Top 50 COVID-19 terms

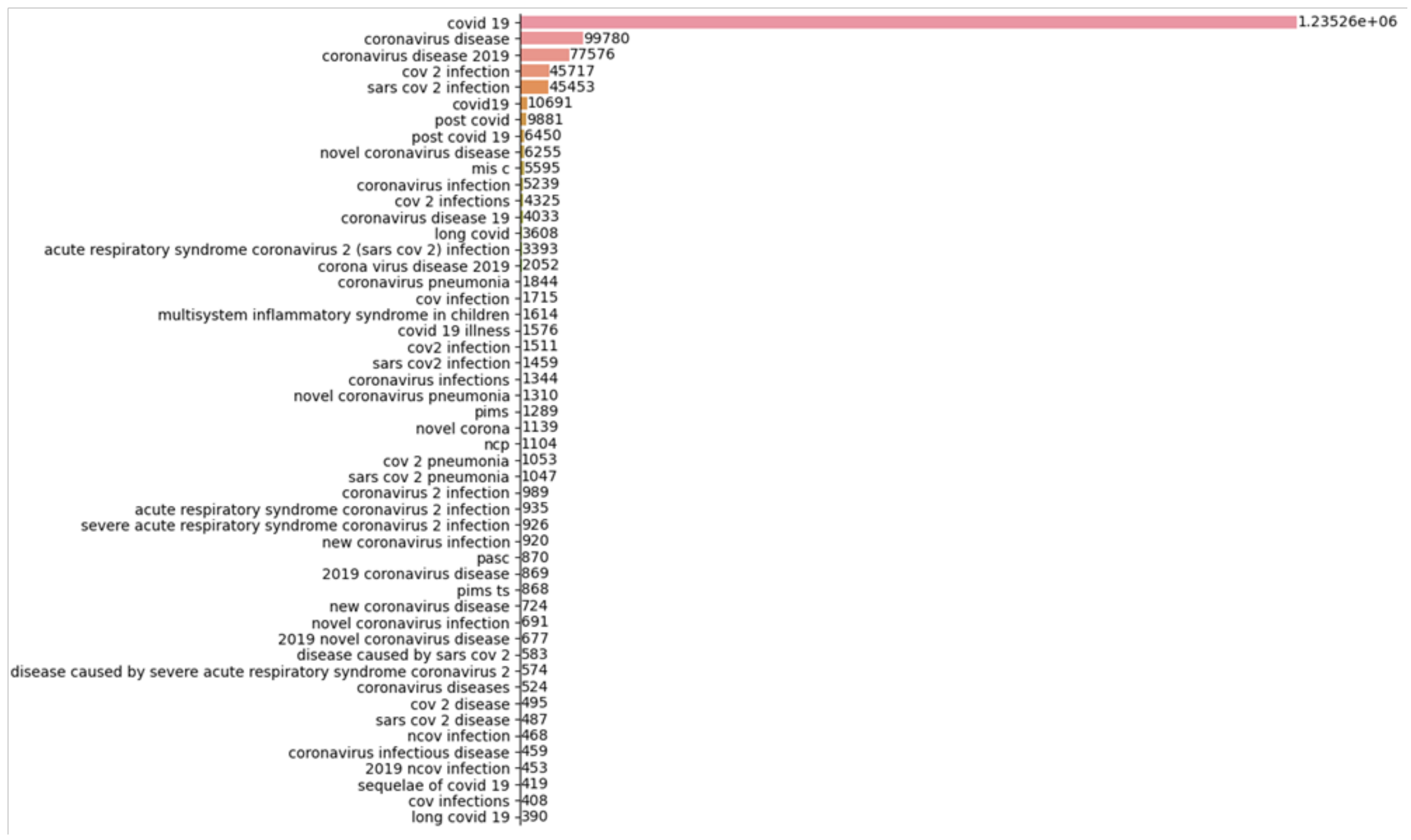

**Figure 1.** Most frequent terms from the SARS-CoV-2 virus and COVID-19 disease dictionaries in abstracts from the CORD-19 dataset. Counts for terms that differed only in hyphenation and casing were aggregated. A) Counts for 50 most frequent virus terms. B) Counts for disease terms. Full results, including all terms with total and document counts, in Supplemental file 11.

To gain insight into the usage of different dictionary terms, we then assessed the frequency of each term, pooling terms that only differed in hyphenation and casing. With the virus dictionary (Figure 1A, Supplemental file 11) the term “sars cov 2”, which corresponds to the official abbreviation SARS-CoV-2 and its spelling variants, was by far the most frequent hit, with 362 422 mentions distributed across ~138 000 abstracts. This was followed quite closely by the general term “coronavirus” (247 173 mentions in ~180 000 abstracts). However, even alternative terms such as “ncov”, “novel coronavirus” and “2019 novel coronavirus” had several thousand mentions. In total, 180 different virus terms were detected after casing and hyphenation variants were pooled.

Similarly to the virus dictionary, the most common term detected with the disease dictionary (Figure 1B, Supplemental file 11) was the official abbreviation and its spelling variants (aggregated as “covid 19”, 1 168 839 mentions. This was followed by matches to the official full disease name. Even in this entity class general terms like “novel coronavirus disease” were found to be common. Terms referring to subforms of the disease (e.g. “post covid”, “post covid 19”, “mis c”, “long covid”) were also frequent. Overall, matches for 686 different disease terms were found.

When running the original variant dictionary (Figure 2A, Supplemental file 11), many false positives were observed from short terms consisting of only one or two letters matching with common words (e.g. a, as, be, by, in) and abbreviations (e.g. kg, mg, ml). We therefore removed these short terms and repeated the annotation and analysis with the modified dictionary (Figure 2B, Supplemental file 11). After this, the most frequent terms observed were “delta”, “alpha” and “omicron”, corresponding to variants which caused major disease waves. 842 different variant terms were found. Terms referring to individual samples in the format of Country/Lab Code-Identifier/Year (e.g. “Switzerland/BS-UHB-42203813/2020”), which made up a large portion of the dictionary, were absent from the CORD-19 abstracts.

For the mutation dictionary (Figure 3, Supplemental file 11), the most common term was “d614g”. This mutation, which arose already early in the pandemic, had over 2000 mentions. Others also still had counts over 1000 (“n5101y) or at least in the hundreds. In total, 56 mutation terms, i.e. half of all dictionary terms, were found.

Overall, the Lund-Annotated-CORD-19 corpus, contains a very large number of texts and annotations. Many different terms were found for each dictionary, reflecting the richness of both the dictionaries and corpus. Nevertheless, a handful of terms dominated the entity counts for each dictionary (Figures 1-3, Supplemental file 11). We also observed large differences in the overall number of detected matches between the dictionaries (Figure 4), with the most common entity class being “virus” followed by “disease”.

Some misidentifications and false-negatives were found, as expected from a dictionary-based annotation method. In addition, even though the CORD-19 dataset was compiled as a literature collection related to COVID-19, many abstracts did not produce matches with the disease and virus dictionary. Manual review of a subset of these abstracts revealed that they often did not directly describe COVID-19 but rather covered related topics, e.g. other pathogens, cellular processes, epidemiology or immunology topics. There was thus no indication, that many articles were missed because of dictionary incompleteness.

Because of licencing restrictions on CORD-19, only part of the Lund-Annotated-CORD-19 corpus, which had permissive licences, can be distributed with this article. However, the full corpus can be assembled with the scripts and instructions in the toolbox and end users can also use them to annotate any other text.

A Top 50 variant terms (all)

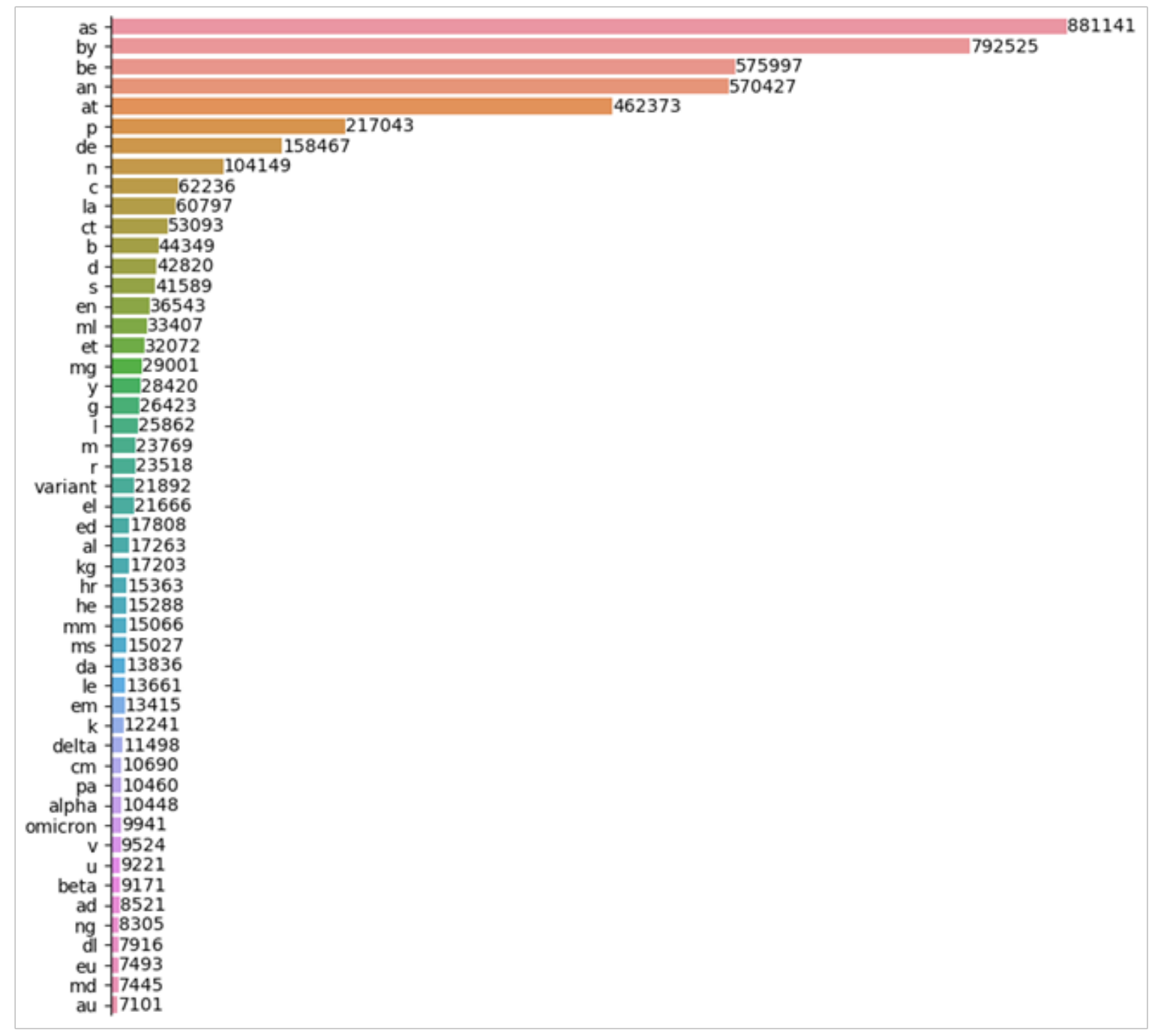

B Top 50 variant terms (≥ 3 characters length)

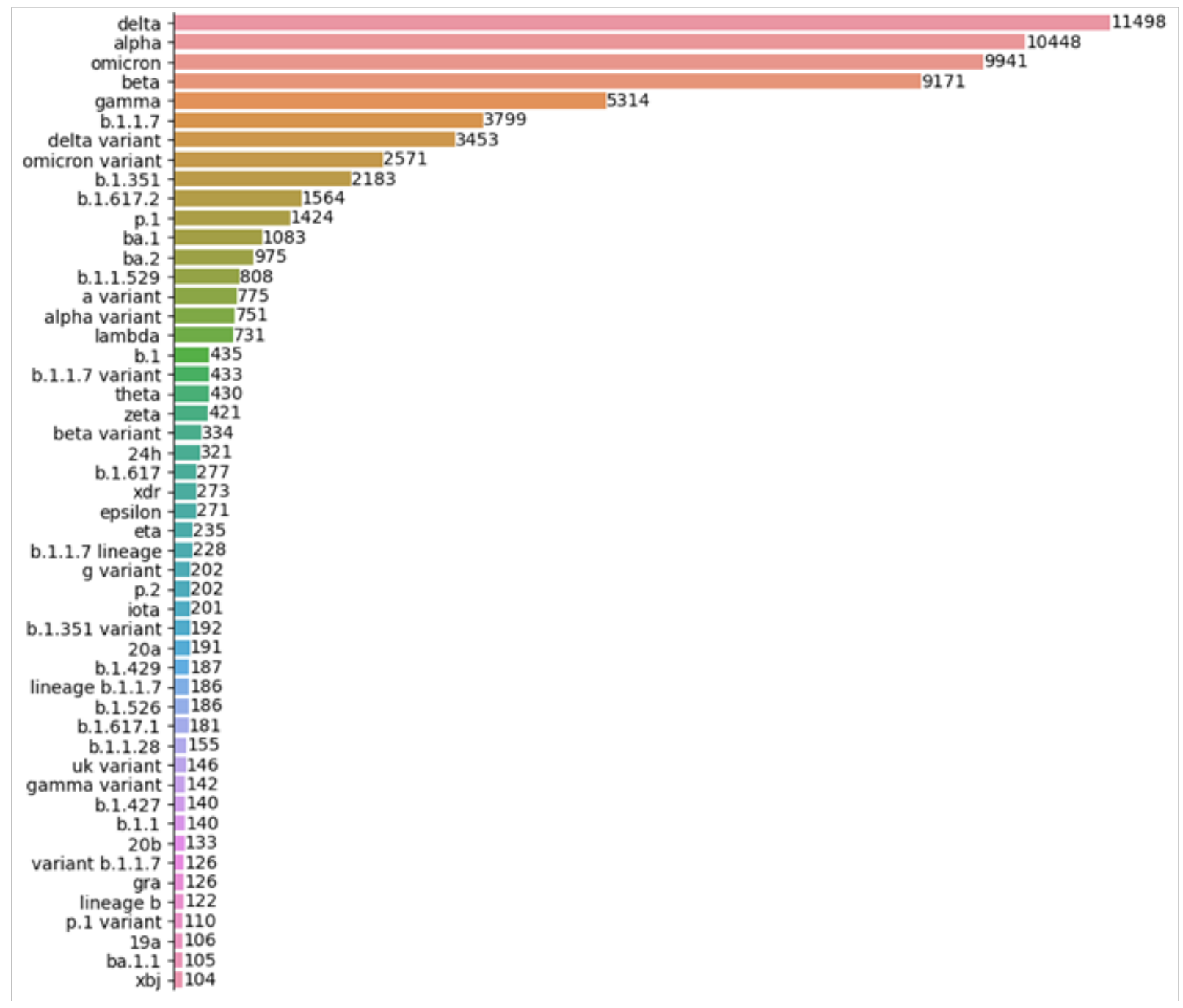

**Figure 2.** Most frequent terms from the SARS-CoV-2 variant dictionary in abstracts from the CORD-19 dataset. Counts for terms that differed only in hyphenation and casing were aggregated. A) Counts for the full dictionary. B) Counts after removal of terms with less than 3 characters. Full results, including all terms with total and per document counts, in Supplemental file 11.

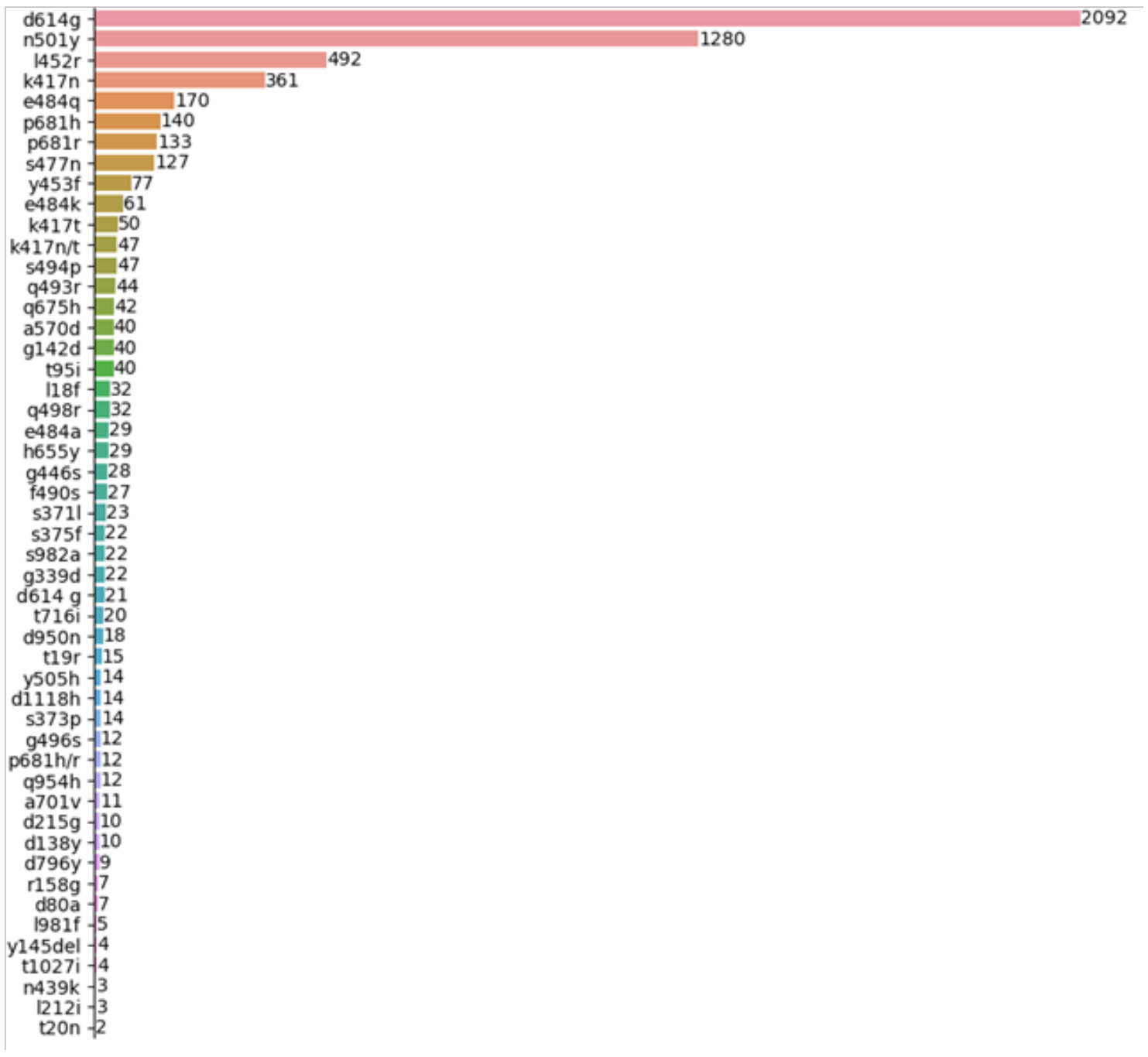

**Figure 3.** Most frequent terms from the SARS-CoV-2 mutation dictionary in abstracts from the CORD-19 dataset. Counts for terms that differed only in hyphenation and casing were aggregated. Full results, including all terms with total and per document counts, in Supplemental file 11.

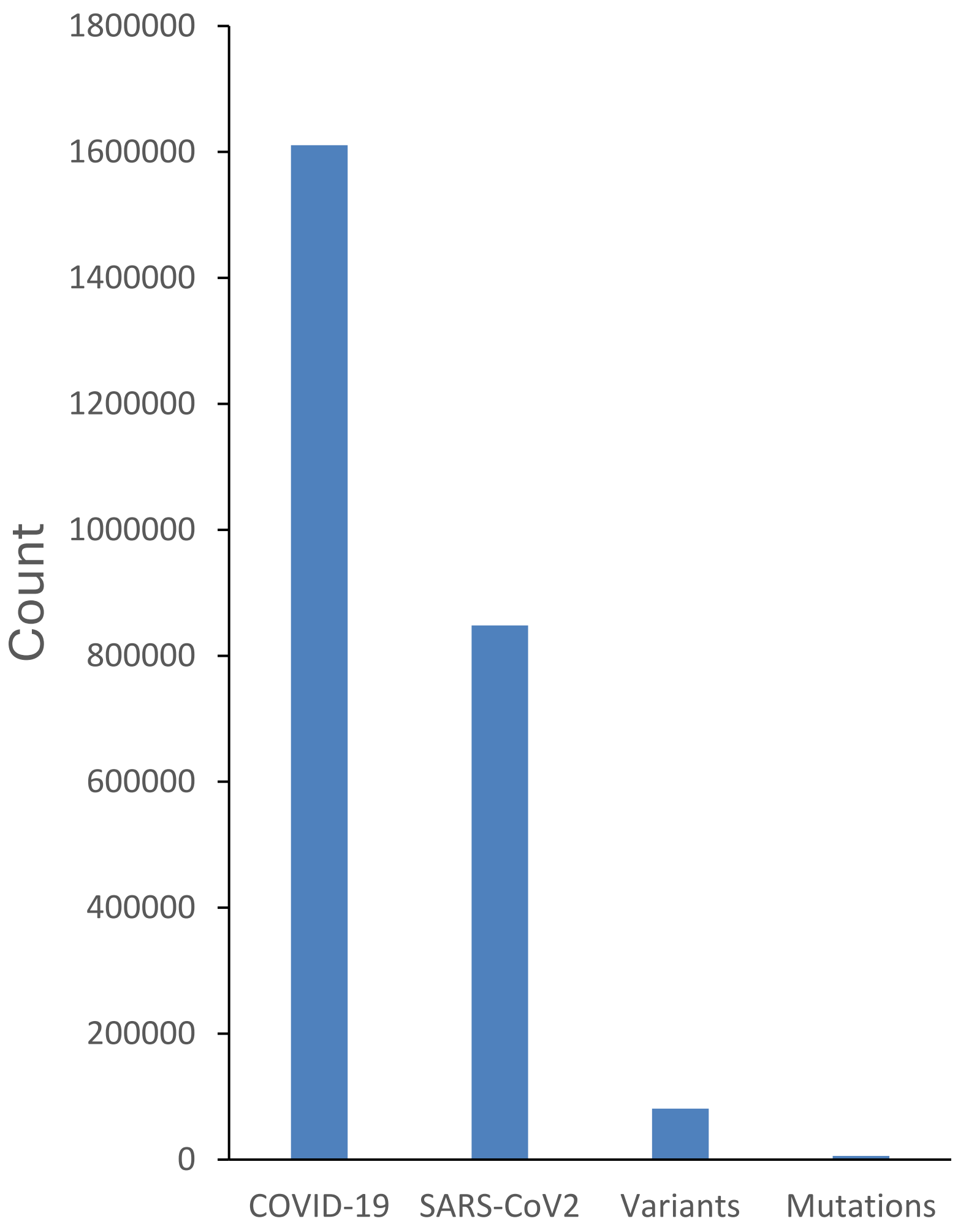

**Figure 4**. Number of matches in CORD-19 across the four dictionaries. Counts for terms that differed only in hyphenation and casing were aggregated. Full results, including all terms with total and per document counts, in Supplemental file 11.

### *Production of the Lund-COVID-19 gold standard corpus*

Small gold standard corpora, with high quality annotations made by domain experts, are needed for thorough testing of NLP tools. Such corpora can also be used for training models with semi-supervised learning, which combines a small amount of labelled data with a large amount of unlabelled data. To produce the Lund-COVID-19 gold standard corpus, 10 randomly selected PubMed abstracts related to SARS-CoV-2/COVID-19 were annotated in BioQRator by the senior author, who is a biomedical expert. Labels were produced for the following classes: cells, chemicals, COVID-19, other diseases, genes/proteins, symptoms, humans, other species, SARS-CoV-2, other viruses and virus families. BioQRator does not allow for overlapping annotations which was problematic when annotating the expression “respiratory, enteric and systemic infections”. In this case, the words respiratory and enteric were annotated as individual entities of the Disease_other category. In total, the Lund COVID-19 corpus (Version 2) contains 199 annotations across the 11 concepts (Table 1). The corpus is available in BioQRator csv, BioC xml and BioCjson format (Supplemental files 8-10).

**Table 1.** Entity counts for the articles in the Corona gold standard corpus. Abstracts are indicated by PubMed ID (PMID).

| PMID | 31986264 | 31991541 | 31992388 | 31996494 | 32007643 | 32013309 | 32015508 | 32020836 | 32029004 | 32036774 | Total |
|---|---|---|---|---|---|---|---|---|---|---|---|
| Cell | | | | | | 5 | | | | | **5** |
| Chemical | | | | | | 1 | | | | | **1** |
| Disease_COVID_19 | 3 | | 2 | | 1 | | 2 | | 6 | | **14** |
| Disease_other | 5 | | | | 3 | 5 | 5 | 3 | 2 | 3 | **26** |
| Genes/Protein | 8 | | | | | 2 | | 4 | | | **14** |
| Symptom | 16 | | | | | 7 | 3 | 2 | | | **28** |
| Species_human | 17 | | | | | 2 | 5 | 3 | | 1 | **28** |
| Species_other | | | | | | | 1 | 2 | | 4 | **7** |
| Virus_family | | 1 | | 3 | | 5 | 3 | 4 | | 3 | **19** |
| Virus_other | 1 | 1 | | | | | | 3 | | 4 | **9** |
| Virus_SARS-CoV-2 | 5 | 4 | 2 | 7 | 6 | 1 | 5 | 5 | 3 | 10 | **48** |
| **Total** | **55** | **6** | **4** | **10** | **10** | **28** | **24** | **26** | **11** | **25** | **199** |

## Discussion

The COVID-19 pandemic has resulted in an explosion of texts which contain valuable information for research, innovation and decision-making both in the medical domain and other areas. This includes, for instance scientific articles, electronic health records, reports, experimental and treatment protocols, policy decisions, news texts and social media posts. Unfortunately, the scale of these texts far exceeds the capacity of human readers. To handle this challenge, automated text processing using NLP approaches is essential. Here we present a COVID-19/SARS-CoV-2 NLP toolbox (Figure 5), containing dictionaries, gold and silver standard corpora, code/workflows and usage instructions. This toolbox can support a broad variety of applications. The toolbox is freely available on GitHub (https://github.com/Aitslab/Covid19) and Zenodo (https://doi.org/10.5281/zenodo.15395348), enabling users to adapt it to their needs. The only limitation is that part of the Lund-Annotated-CORD-19 corpus must be generated by the end user themselves, due to licence limitations on the CORD-19 dataset. However, the openly shared part of the corpus is sufficient large for many purposes and re-creating the full corpus is relatively simple by following the instructions and scripts we provided.

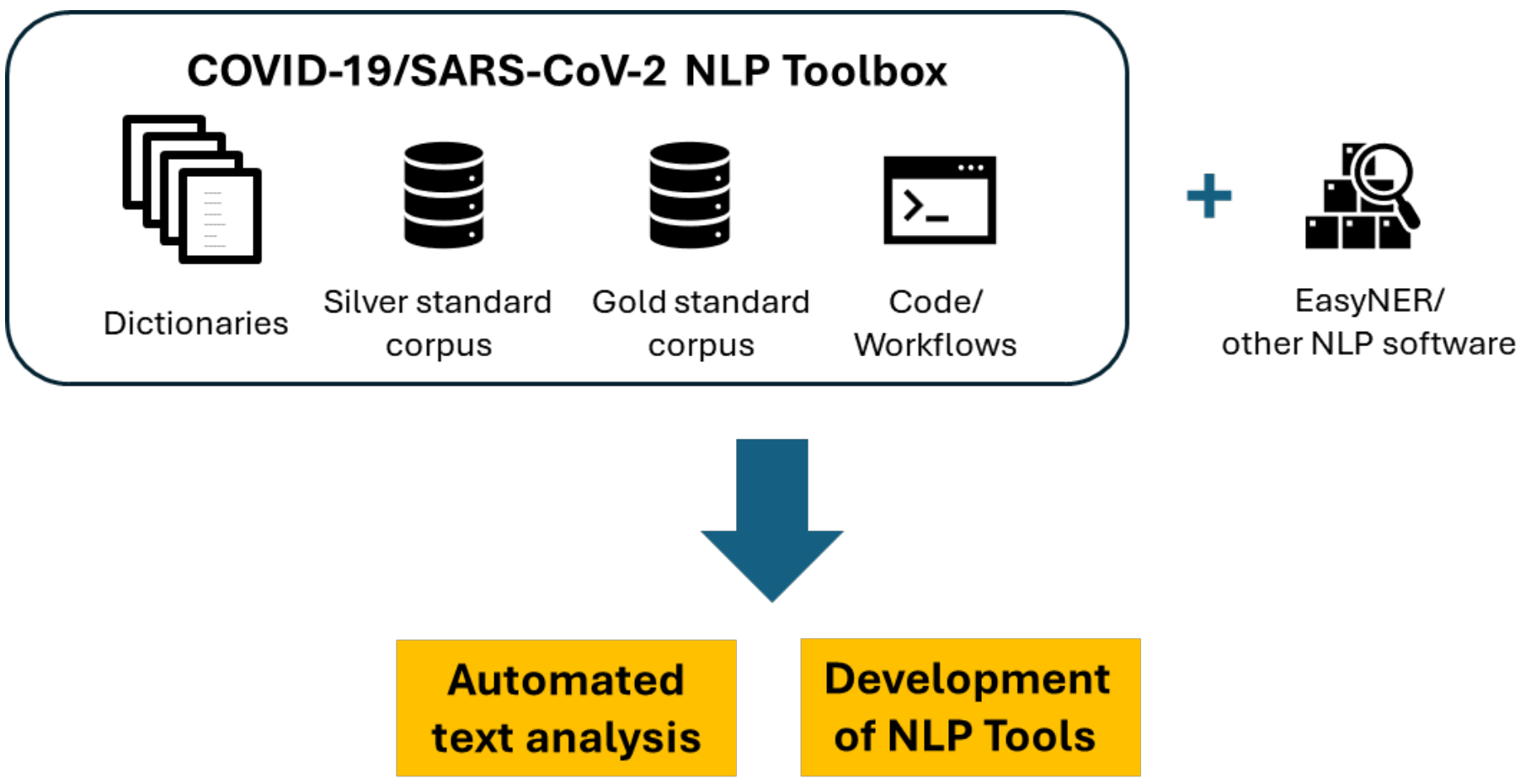

**Figure 5.** Overview over the COVD-19/SARS-CoV-2 NLP Toolbox and its applications.

The toolbox seamlessly integrates with EasyNER [24], an easy-to-use highly customizable information extraction tool developed by our group. However, our toolbox can also be combined with other COVID-19-related BioNLP resources [4] and with subject-agnostic NLP tools such as spaCy [28]. To facilitate this, the dictionaries are formatted as text files with one term per row, which is a standard format used by many tools. The dictionary files can be loaded without modification into EasyNER software [24, 25], and we made use of this capability and various modules of the software to produce and analyse the silver standard corpus. Users can follow this workflow to annotate and analyse their own text collections.

The dictionaries in the toolbox respectively contain COVID-19 disease synonyms, SARS-CoV-2 virus synonyms, as well as synonyms for variants/lineages and common mutations. These dictionaries were created using a combination of manual curation and computational processing. Some of the automatically generated terms are unlikely to be in use but as they would simply be unmatched in NER procedures there is no large benefit from excluding them through time-consuming manual review unless working in a low resource environment. Similarly, some terms such as those referring to individual samples in the variant dictionary, may be rare or absent in some text types, as seen with the CORD-19 dataset, and can be excluded by the end user to save memory and reduce processing time if needed.

To minimize dictionary size, we removed capitalization and hyphenation. EasyNER can already detect entities regardless of casing and we provided additional scripts for pooling entity counts for hyphenation and casing variants. When using other tools, users need to ensure that appropriate measures for preprocessing or incomplete matching (so-called "fuzzy" matching) are implemented. Fuzzy matching can also help detect plural variants but as this is relatively easy to incorporate and mostly relevant for a few disease terms such as "coronavirus infections" we did not implement it here.

One of the major limitations of dictionary-based matching is the risk of false positives from terms that have multiple meanings. We therefore excluded some highly ambiguous terms such as those with only 1-2 characters. If needed, users can update the dictionaries and silver standard corpus easily, both by manually writing terms into the files or by using the scripts and instructions for dictionary production in the toolbox. The same scripts also allow for periodic updates of the dictionaries, for example to include emerging mutations and variants.

Deep learning-based NER, can improve NER performance by taking context into account and our toolbox can easily be leveraged to develop such models using supervised and semi-supervised approaches. Making use of our dictionaries and EasyNER's in-built support for processing CORD-19, we already created a very large silver standard corpus, which can be used for training and evaluation deep leaning models, alone or in combination with the provided gold standard corpus. However, model development requires more resources and for many applications NER with our dictionaries is sufficient.

Recently, the first version of our toolbox has been used in a framework to generate a comprehensive COVID-19 knowledge graph [23, 29]. Such knowledge graphs can be an important tool to direct experimental work or support systematic literature review. In addition, our toolbox has been used in a linguistic study on COVID-19 terminology [6]. Other potential uses include information extraction from electronic health records for clinical research, analysis of social media and news items for social and political science applications, and many other areas where large amounts of COVID-19-related text need to be processed.

Even though the toolbox is designed for the English language, the dictionaries can also be applied for NLP in other languages because many of the terms are highly similar across different languages. For such applications, small language differences can be handled using incomplete matching approaches and additional language-specific terms can be added manually or with the help of computational translation tools.

In addition to generating the toolbox, we applied it together with EasyNER to perform an in-depth analysis of the CORD-19-dataset, which was used to produce the silver standard corpus. This is highly relevant, as CORD-19 is one of the most comprehensive collections of COVID-19-related research literature. Our analysis revealed that hundreds of synonymous terms for SARS-CoV2 and COVID-19 are present in research literature produced during the pandemic, confirming findings from other studies [6]. Furthermore, we observed a significant amount of CORD-19 abstracts that did not match terms from the virus and disease dictionaries. There was no indication that the dictionaries were missing important terms. Instead, many of these abstracts did not directly describe COVID-19/SARS-CoV-2 but were included in CORD-19 as they describe related topics and can contain information that is highly relevant to COVID-19 researchers and others working with the disease. Both the high degree of lexical variation and the existence of relevant information in articles not directly describing COVID-19/SARS-CoV-2, need to be taken into account when searching for documents and extracting information by manual or computational methods.
Furthermore, our analysis revealed that there are hundreds or even thousands of articles which describe individual SARS-CoV-2 mutations and variants. Integrating this knowledge will be paramount to better understand the disease, which still has a significant impact worldwide. Despite the richness of the COVID-19 research literature our analysis also revealed significant imbalance in the CORD-19 dataset. It was not surprising that the official names and abbreviations, accounted for the majority of disease and variant mentions. However, a similar imbalance was seen for mutations and variants, with a few terms dominating the research literature. This reflects real gaps and biases in the research and our analysis can thus be used to highlight mutations and variants in need of more research.

**Conclusion**

The presented toolbox enables the analysis of text collections related to SARS-CoV2/COVID-19 on a large scale and integrates well with existing resources, in particularly EasyNER. The included dictionaries and corpora can also easily be updated and expanded using the provided scripts and instructions and users are very welcome to contribute to the continued development of this open community resource. Besides its direct use for information extraction, the toolbox can be applied in the development of other NLP tools, such as deep learning models.
Applying the toolbox to CORD-19, a very large collection of research articles related to SARS-CoV2/COVID-19, revealed a very large lexical variation of terms and large differences in the frequency of different key entities. This needs to be taken into account when working with NLP approaches in this domain.

**Acknowledgement and funding**

The computations and data handling were enabled by resources provided by the National Academic Infrastructure for Supercomputing in Sweden (NAISS) at Lund University (LUNARC).
We thank the following funders who support our research: the SciLifeLab/Knut and Alice Wallenberg Foundation National COVID-19 Research Program, the Swedish Research Council, the Swedish Research Council for Sustainable Development (FORMAS), the Segerfalk Foundation, the Swedish Brain Foundation, the Crafoord Foundation and the Royal Physiographic Society.
We also acknowledge the following research environments and networks which support our work: AI Lund, AIR Lund, Lund University Profile Area “Nature-based Future Solutions”, Lund University Profile Area “Natural and Artificial Cognition”, Lund University Profile Area “Proactive Ageing”, LTH Profile Area “AI and Digitalization”, LTH Profile Area “Engineering Health”, Strategic Research Area “EpiHealth”, Strategic Research Area “eSSENCE”, and PhenoTarget. We thank Swe-Clarin for support, the NLM/NCBI BioNLP Research Group for making their BioC-JSON tool and term lists publicly available, and Pierre Nugues and Marcus Klang from the Department of Computer Science at Lund University for valuable advice throughout the project.

**List of abbreviations**

| | |
|---|---|
| NLP | Natural Language Processing |
| NER | Named Entity Recognition |

**Competing Interests**

We have no competing interest in connection with this publication.

**Authors' contributions**

Conceptualization/Methodology: SA, SKR, JF; Data curation/Formal analysis/Investigation/Software/Validation: SA, SKR, RA; Funding acquisition/Project administration/Supervision: SA; Writing – original draft: SA, SKR; Writing – review & editing: SA, SKR, RA

**Availability of data and materials**

The datasets and scripts of this article are available in the Zenodo repository and on GitHub (see below)

Project name: SARS-CoV-2 NLP toolbox
Project home page: https://github.com/Aitslab/Covid19
DOI: https://doi.org/10.5281/zenodo.15395348
Operating system: Platform independent
Programming language: Python
Other requirements: Python v3.6 or higher, PyBioC (https://github.com/2mh/PyBioC), pandas v1.4.0 or higher, spacy v2.2
Licence: code and corpus annotations – Apache 2.0; CORD-19 abstract texts as in original dataset (see metadata.csv file)

**Supplemental material**

Supplemental file 1
SARS-CoV-2 dictionary, version 3

Supplemental file 2
COVID-19 dictionary, version 3

Supplemental file 3
SARS-CoV-2 variant dictionary, version 2 (including terms with 1 and 2 characters)

Supplemental file 4
SARS-CoV-2 protein mutation dictionary, version 1

Supplemental file 5

Zip file with scripts

Supplemental file 6
Zip file with the partial Lund-Annotated-CORD-19 corpus, and instructions for obtaining the full corpus

Supplemental file 7
SARS-CoV-2 variant dictionary, version 2 (excluding terms with 1 and 2 characters)

Supplemental file 8
Lund-COVID-19 gold standard corpus, version 2 in BioC xml format

Supplemental file 9
Lund-COVID-19 gold standard corpus, version 2 in BioC json format

Supplemental file 10
Lund-COVID-19 gold standard corpus, version 2 in BioQRator csv format

Supplemental file 11
Zip file with entity counts for the Lund-Annotated-CORD-19 corpus

**References**

1. Masters-Waage TC, Jha N, Reb J: **COVID-19, Coronavirus, Wuhan Virus, or China Virus? Understanding How to "Do No Harm" When Naming an Infectious Disease**. *Front Psychol* 2020, **11**:561270.
2. Santoni D, Vergni D: **In the search of potential epitopes for Wuhan seafood market pneumonia virus using high order nullomers**. *J Immunol Methods* 2020, **481-482**:112787.
3. Cook HV, Jensen LJ: **A Guide to Dictionary-Based Text Mining**. *Methods Mol Biol* 2019, **1939**:73-89.
4. Peng J, Xu D, Lee R, Xu S, Zhou Y, Wang K: **Expediting knowledge acquisition by a web framework for Knowledge Graph Exploration and Visualization (KGEV): case studies on COVID-19 and Human Phenotype Ontology**. *BMC Med Inform Decis Mak* 2022, **22**(Suppl 2):147.
5. Anderson BS: **Using text mining to glean insights from COVID-19 literature**. *J Inf Sci* 2023, **49**(2):373-381.
6. Leaman R, Lu Z: **A Comprehensive Dictionary and Term Variation Analysis for COVID-19 and SARS-CoV-2**. In*: dec 2020; Online*. Association for Computational Linguistics.
7. Kazemi Rashed S, Ahmed R, Frid J, Aits S. **Files and code for English dictionaries, gold and silver standard corpora for biomedical natural language processing related to SARS-CoV-2 and COVID-19**
8. Federhen S: **The NCBI Taxonomy database**. *Nucleic Acids Res* 2012, **40**(Database issue):D136-143.
9. WHO. **International Classification of Diseases, v10** https://icd.who.int/browse10/2019/en
10. WHO. **International Classification of Diseases, v11** https://icd.who.int/browse11
11. Schriml LM, Mitraka E, Munro J, Tauber B, Schor M, Nickle L, Felix V, Jeng L, Bearer

C, Lichenstein R *et al*: **Human Disease Ontology 2018 update: classification, content and workflow expansion**. *Nucleic Acids Res* 2019, **47**(D1):D955-D962.
12. National Library of Medicine. **Medical Subject Headings (MeSH)** https://www.ncbi.nlm.nih.gov/mesh/),
13. **COVID-19 and SARS-CoV-2 term variation** https://github.com/ncbi-nlp/CovidTermVar
14. WHO. **Tracking SARS-CoV-2 variants** https://www.who.int/en/activities/tracking-SARS-CoV-2-variants/
15. Shu Y, McCauley J: **GISAID: Global initiative on sharing all influenza data - from vision to reality**. *Euro Surveill* 2017, **22**(13).
16. Hadfield J, Megill C, Bell SM, Huddleston J, Potter B, Callender C, Sagulenko P, Bedford T, Neher RA: **Nextstrain: real-time tracking of pathogen evolution**. *Bioinformatics* 2018, **34**(23):4121-4123.
17. Nextstrain. **Nextstrain Clades** https://github.com/nextstrain/ncov/blob/master/defaults/clades.tsv
18. **Pango Lineages** https://cov-lineages.org/lineage_list.html
19. **Pango Taxon List** https://raw.githubusercontent.com/cov-lineages/pango-designation/refs/heads/master/lineages.csv
20. **Pango Alias File** https://github.com/cov-lineages/pango-designation/blob/master/pango_designation/alias_key.json
21. Sarkar M, Madabhavi I: **COVID-19 mutations: An overview**. *World J Methodol* 2024, **14**(3):89761.
22. Farhud DD, Mojahed N: **SARS-COV-2 Notable Mutations and Variants: A Review Article**. *Iran J Public Health* 2022, **51**(7):1494-1501.
23. Wang LL, Lo K, Chandrasekhar Y, Reas R, Yang J, Burdick D, Eide D, Funk K, Katsis Y, Kinney R *et al*: **CORD-19: The COVID-19 Open Research Dataset**. *ArXiv* 2020.
24. Ahmed R, Berntsson P, Skafte A, Rashed SK, Klang M, Barvesten A, Olde O, Lindholm W, Arrizabalaga AL, Nugues P *et al*: **EasyNER: A Customizable Easy-to-Use Pipeline for Deep Learning- and Dictionary-based Named Entity Recognition from Medical Text**. *arxiv* 2023.
25. Aits S, Ahmed R. **EasyNER** https://github.com/Aitslab/EasyNER
26. **CORD-19** https://github.com/allenai/cord19
27. **BioC-JSON conversion tool**. In.
28. Honnibal M, Montani I, Van Landeghem S, Boyd A: **spaCy: Industrial-strength Natural Language Processing in Python**. 2020.
29. Chen Q, Allot A, Lu Z: **Keep up with the latest coronavirus research**. *Nature* 2020, **579**(7798):193.